\documentclass{ws-ijmpa}
\begin{document}
\title{MESON PRODUCTION AND BARYON RESONANCES AT CLAS
\footnote{Invited talk given at MESON 2010, Krakow, Poland.}}
\author{VOLKER D. BURKERT} 
\address{Jefferson Lab, 12000 Jefferson Avenue, 12000 Jefferson \\
Newport News, Virginia, 26000, USA \\ burkert@jlab.org}
\maketitle

\begin{abstract}
I give a brief overview of the exploration of baryon properties in meson photo- and electroproduction. These processes provide ample information for the study of electromagnetic couplings of baryon resonances and to search for states, yet to be discovered. The CLAS detector, combined with the use of energy-tagged polarized photons and polarized electrons, as well as polarized targets and the measurement of recoil polarization, provide the tools for a comprehensive nucleon resonance program. I briefly present the status of this program, prospects for the next few years, and plans for the Jefferson Lab 12 GeV upgrade.
\end{abstract}
\ccode {PACS numbers 1.55Fv, 13.60Le, 13.40Gp, 14.20Gk}

\section{Introduction}	
The systematics of the baryon excitation spectrum is key to understanding the effective degrees of freedom underlying 
the nucleon matter\cite{burkert-lee}. The most comprehensive predictions 
of the resonance excitation spectrum come from the various implementation 
of the constituent quark model based on broken $SU(6)$ 
symmetry\cite{isgur}. Gluonic degrees of freedom may also play a role\cite{libuli}, and resonances may  
be generated dynamically through  
baryon-meson interactions\cite{oset}. Recent developments in  Lattice QCD led to predictions
 of the nucleon spectrum in QCD with dynamical quarks\cite{dgr}, albeit with still large pion 
 masses of ~420 MeV.  In parallel, the development of dynamical coupled 
 channel models is being advanced with new vigor. The EBAC group at JLab has
 demonstrated\cite{kamano} that coupled channel effects result in a large mass shift for the Roper resonance downward
 from the bare core mass of ~1736 MeV to ~1365 MeV, explaining the low physical mass of the state.
 
The various resonance models not only 
predict different excitation spectra but also different $Q^2$ dependence of 
transition form factors.  Mapping out transition form factors in meson electroproduction 
experiments can tell us a great deal about the effective degrees of freedom underlying 
baryon structure.    

\section{\bf Search for new excited nucleon states}    
It is well recognized that the analysis of differential cross sections in the
photoproduction of single pseudoscalar mesons alone results in ambiguous solutions 
for the contributing resonant partial waves. The $N^*$ program at JLab is  
aimed at complete, or nearly complete measurements for processes such as
$\vec{\gamma} \vec{p} \rightarrow \pi N, ~\eta p,~K^+\vec{Y}$ and 
$\vec{\gamma} \vec{n} \rightarrow \pi N, ~K^\circ\vec\Lambda$. These 
reactions are fully described by four complex
 parity-conserving amplitudes, which may be determined from eight well-chosen combinations of 
unpolarized cross sections, and single and double polarization observables using beam, 
target, and recoil polarization measurements. If all combinations are 
measured, 16 observables can be extracted providing highly redundant information for the 
determination of production amplitudes. 

A large part of the experimental program makes use of the CLAS detector\cite{clas}, 
which provides particle identification and
momentum analysis in a polar angle range from 8$^\circ$ to 140$^\circ$. The 
photon energy tagger provides energy-marked photons with an energy resolution of 
${\sigma(E) \over E} = 10^{-3}$. Circularly polarized photons are generated by scattering 
the highly polarized electron beam from an amorphous radiator. Other equipment includes a coherent 
bremsstrahlung facility with a precision goniometer for diamond crystal positioning and 
angle control. The facility has been used for coherent photon bremsstrahlung production, 
generating photons with linear polarization up to 90\%. There are two frozen spin polarized 
targets, one based on butanol as target material (FROST), and one using frozen HD as 
target material (HDIce). FROST has been operated successfully. An production run of 5 months 
has been completed in August 2010. The HDIce target is currently being assembled. It has 
better dilution factor and will serve as polarized neutron target. It is scheduled to take data in 2011 and 2012. 
\begin{figure}[t]

\begin{center}
\includegraphics[width=12cm,height=10cm]{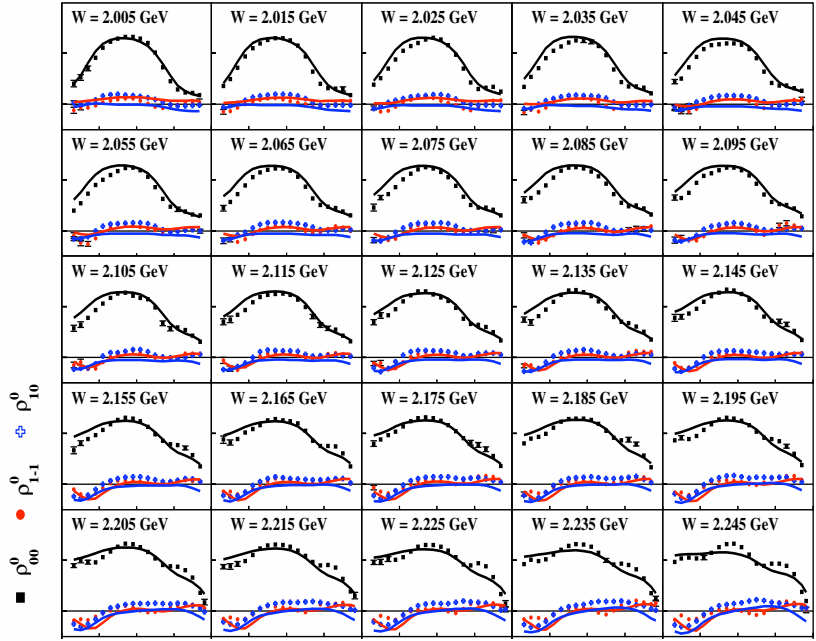}
\caption{Angular dependence in the range $-1 < cos\theta^* < 1$ for spin density matrix elements $\rho_{00}^0$, $\rho_{10}^0$, and $\rho_{1-1}^0$ of the process $\gamma p \to p \omega$, in the mass range from 2 to 2.25 GeV.}
\label{fig:omega_spin_density}
\end{center}
\end{figure} 

\begin{figure}[t]
\begin{center}
\includegraphics[width=12cm, height=8cm]{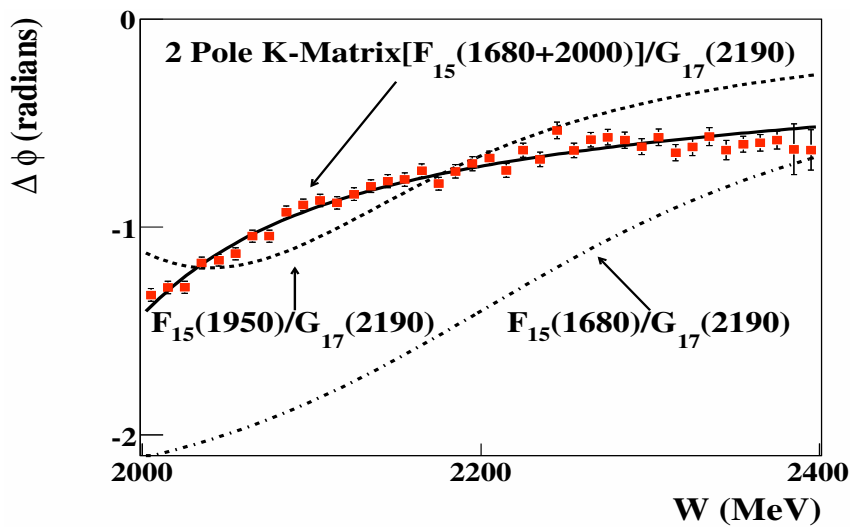}
\caption{Fit results for $\Delta\phi=\phi_{7/2^-}-\phi_{5/2^+}$  vs 
the invariant $p\omega$ mass. The dot-dashed line is the phase motion expected 
 using constant width Breit-Wigner distributions and the parameters quoted by the PDG for the
  $F_{15}(1680)$ and $G_{17}(2190)$. The dashed line required the $J^P =7/2^-$ parameters to be 
within the PDG limits for the $G_{17}(2190)$, while allowing the $J^P = 5/2^+$ parameters to vary freely. 
The solid line used a constant width Breit-Wigner distribution for the $G_{17}(2190)$, 
but a 2-pole single channel K-matrix for the $J^P=5/2^+$ wave.}
\label{fig:omega_pwa}
\end{center}
\end{figure}

In hyperon channels, very precise cross section and polarization data have been 
measured\cite{mcnabb04,brad,hleiqawi,mccrack10,pereira10,dey10,guo07}. In channels involving nucleons 
in the final state where the recoil polarization is usually not measured, seven independent observables 
are obtained directly. The recoil polarization can be inferred from a beam-target double polarization measurement.  
Precise measurements of differential cross section for $\pi^\circ$, $\pi^+$, $\pi^-$ $\eta$,
$\eta^{\prime}$, and $\omega$ have been 
published\cite{dugger,wchen,mwilliams}, while 
analyses of polarization observables for these reactions are in progress. For $p\omega$ 
production,  the $\omega \to \pi^+\pi^-\pi^{\circ}$ decay distribution contains polarization information, providing 
additional information that constrain the partial wave analysis. In CLAS, the final state is fully determined by measuring
the charged pions, and inferring the $\pi^\circ$ through kinematical constraints.  
Fig.~\ref{fig:omega_spin_density} shows the spin density matrix 
elements for $\gamma p \to p \omega$. Results of the partial wave fit are shown in Fig.~\ref{fig:omega_pwa}. 
The fit to the phase motion of dominant partial waves requires inclusion of 3 well known resonances, 
$F_{15}(1680),~ D_{13}(1700),~ G_{17}(2190)$), as well as the  quark model state $F_{15}(2000)$, 
an unconfirmed 2-star state in RPP\cite{pdg2008}. 

\section{Electromagnetic excitation of nucleon resonances}
Electromagnetic transition form factors encode information about the transition 
charge and magnetization densities\cite{vdh08}, which in turn reflect the electromagnetic structure of the 
excited states. The non-relativistic constituent quark model (nrCQM) provides a reasonable 
representation of the mass spectrum of many excited 
states below 2 GeV. Relativized versions, e.g. in ref. [\refcite{santopinto}], give 
qualitative agreement with some of the measured transition amplitudes.  
The $N(1440)P_{11}$, or  "Roper" resonance, however, has escaped description 
within this approach. The constituent quark model represents this state 
as a radial excitation of the nucleon, but has difficulties to describe its basic
features such as the mass, photocouplings, and $Q^2$ evolution. In fact, the photocoupling 
amplitude has the wrong sign, and its magnitude is predicted to rise strongly at small $Q^2$ contrary 
to the data that show a rapid drop. 

\subsection{The Roper resonance}
 
 The problems with describing the $N(1440)P_{11}$ "Roper" in constituent quark models has 
prompted the development of alternative models involving gluon 
fields\cite{libuli}, or meson-baryon degrees of freedom\cite{cano,krewald}, 
and light-cone quark model\cite{capstick,aznauryan-qm}. 
In particular, measurements of its transition amplitudes in pion electroproduction has revealed 
information about the nature of the state.  

\begin{figure}[t]
\begin{center}
\includegraphics[width=12cm,height=6.cm]{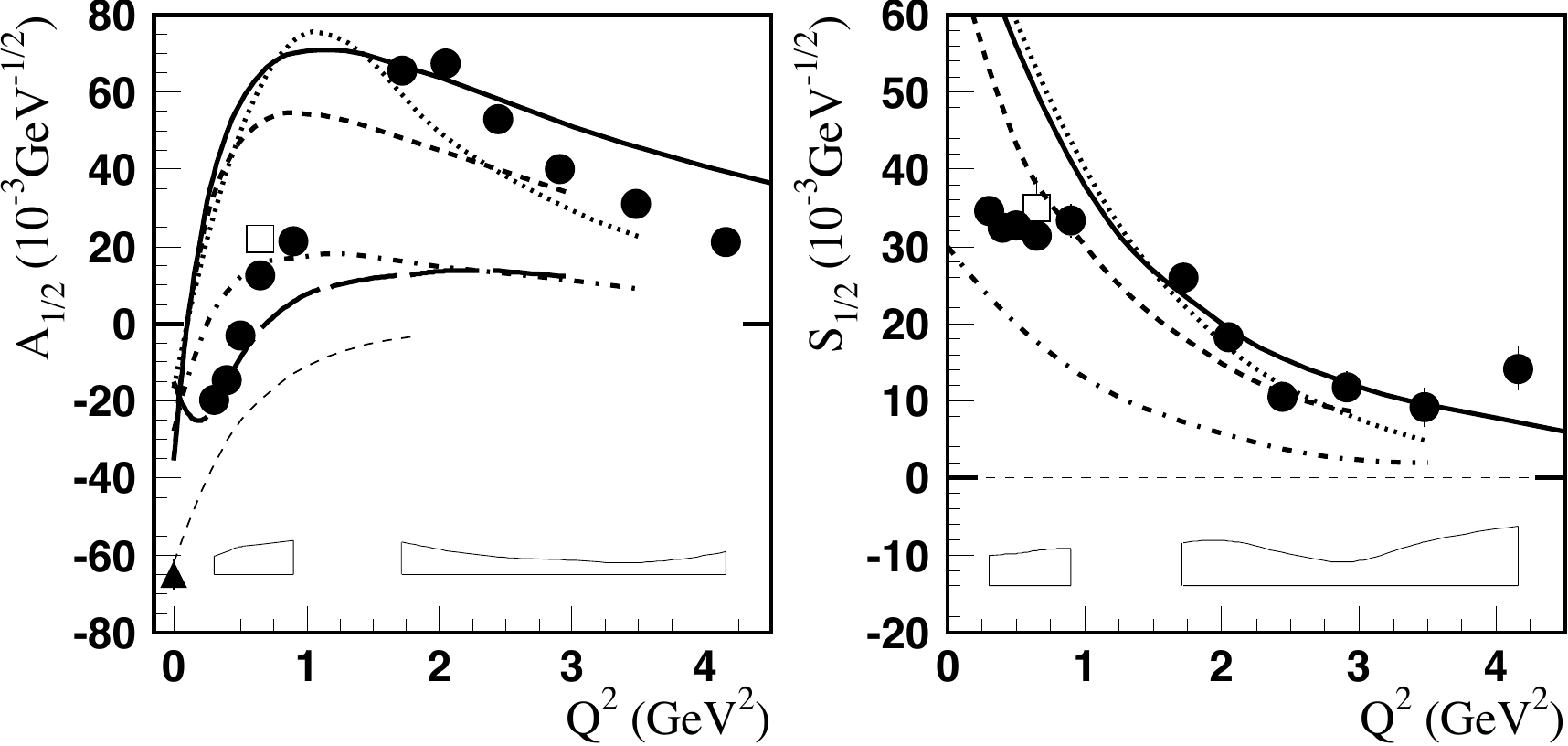}
\caption {Transverse electrocoupling amplitude for the 
Roper $N(1440)P_{11}$ (left panel). The full circles 
are the new CLAS results. The squares are previously published results of fits to CLAS data
at low $Q^2$. The right panel shows the longitudinal amplitude. The bold curves are all 
relativistic light front quark model calculations \protect\cite{aznauryan-qm}. The thin dashed 
line is for a gluonic excitation\protect\cite{libuli}. } 
\end{center}
\label{roper}
\end{figure}

Given these different theoretical concepts for the structure of the state, 
the question ``what is the nature of the Roper state?'' has been a focus of 
the $N^*$ program with CLAS.  The state is very wide, and pion electroproduction 
data covering a large range in the 
invariant mass W with full center-of-mass angular coverage are key in extracting the 
transition form factors. As an isospin  $I = {1\over 2}$ state, 
the $P_{11}(1440)$ couples more strongly to n$\pi^+$ than to p$\pi^\circ$. Also,
contributions of the high energy tail of the $\Delta(1232)$ are much reduced in that 
channel due to the $I = {3\over 2}$ nature of the $\Delta(1232)$. Previous 
studies\cite{maid07} have mostly used the $p\pi^0$ final state from 
measurements focussing on the $\Delta(1232)$ mass region. Recently published analysis 
by the CLAS collaboration included new high statistics $n\pi^+$ data\cite{park08} that covered the 
mass region from pion threshold to $1.7$~GeV/c$^2$. In addition, 
large samples of differential cross section data\cite{kjoo,egiyan06,ungaro06}  and polarization 
observables\cite{biselli}  taken earlier with CLAS 
in the channels $p\pi^\circ$ and $n\pi^+$ are
included in a comprehensive analysis of nearly 120,000 data points covering a large range 
in $\theta_{\pi},~\phi_\pi,~W,~Q^2$. Also, cross section data from $p\eta$ final state\cite{thompson01,denizli07}
were used to independently determine amplitudes of the transition to the $N(1535)S_{11}$ state. This 
allows one to constrain the branching ratio $\beta_{N\pi}$ and $\beta_{N\eta}$ for this state more accurately.
Two independent approaches, fixed-t dispersion relations
and the unitary isobar model, have been employed to estimate the model-sensitivity of the
resulting transition amplitudes\cite{aznauryan08,aznauryan09} for the three low mass states,
$N(1440)P_{11}$, $N(1520)D_{13}$, and $N(1535)S_{11}$. 

The transverse and longitudinal amplitudes $A_{1/2}$ and $S_{1/2}$ of the transition to the
 $N(1440)P_{11}$ resonance are shown in Fig.~\ref{roper}.      
At the real photon point $A_{1/2}$ is negative, rises quickly with $Q^2$,  and changes sign 
near $Q^2=0.5$~GeV$^2$. At $Q^2=2$GeV$^2$ the amplitude reaches about the same
 magnitude but opposite sign as at $Q^2=0$, before it slowly falls off with $Q^2$. 
 This remarkable sign change with $Q^2$ has not been observed before for any nucleon form factor or transition 
amplitude.  At high $Q^2$, both amplitudes are qualitatively described by the 
light front quark models, which is consistent with the interpretation of the state as a radial excitation 
of the nucleon at short distances. The low $Q^2$ behavior is not well described by the LF quark models  
which fall short of describing the amplitude at the photon point. This suggests that important
contributions, e.g. meson-baryon interactions describing the large distances behavior, 
are missing, as has also been recently demonstrated in a covariant valence quark model\cite{ramalho,gross}. 
A first exploration of the Roper transition form factors has recently been undertaken within Lattice QCD\cite{roper-lqcd}.

\subsection{Strangeness electroproduction}
Electroproduction cross section measurements of strangeness channel, e.g. $ep\to eK^+Y^\circ$ , 
and polarization transfer measurements $\vec{e}p\to eK^+ \vec{Y^\circ}$, ( $Y^\circ= \Lambda, ~\Sigma$)
have been carried out\cite{ambroz,carman} that allow stringent tests of hadronic models. The separated structure 
functions reveal clear differences between the production dynamics for the $\Lambda$ and $\Sigma$. No hadronic 
model has been able to reproduce the energy and angle dependences, which show indications of strong s-channel resonance
behavior, especially in the case of  the $K^+\Sigma^\circ$ channel.  These data will provide 
important constraints for dynamically coupled-channel analyses.     

\section{Conclusion \& Outlook} 
A large effort is currently underway with CLAS at Jefferson Lab to probe the S=0 baryon excitation spectrum in 
photo- and electroproduction 
measurements of varies baryon-meson final states. These include cross sections and many polarization observables,
both with polarized beam and polarized targets, as well as recoil polarization measurements. Measurement on 
polarized proton targets have been completed with an extended run in 2010. 
The program on the neutron is planned to be completed with a photon run using polarized HD material in 2011/2012.
The determination of transition amplitudes for many high-mass states is ongoing. 
Precise transition amplitudes have been published for the
low-mass resonances. The 2-pion final state is being analyzed to study both low mass states\cite{fedotov} and 
higher mass states that dominantly couple to 2-pion channels such as $S_{31}(1620)$, $D_{33}(1700)$ and 
others\cite{ripani,mokeev}.  Lastly, a program to measure resonance transition form factors at 
$5 < Q^2 <12$~GeV$^2$ has been approved for the CLAS12 detector\cite{burkert} currently under construction at JLab as 
part of the 12 GeV energy upgrade of its electron accelerator.

\end{document}